\documentclass[12pt,a4paper]{panl}
\usepackage{cite}
\usepackage{wrapfig}
\usepackage{graphicx}
\usepackage{amssymb}
\usepackage{amsfonts}
\usepackage{amsmath}
\usepackage{longtable}
\usepackage{rotating}
\usepackage{lscape}
\usepackage{epsfig}
\usepackage{multirow}

\originalTeX
\begin{document}

\title{
Relations between spin observables of the reactions $dd\to  npd$  and  $pd\to pd$ in impulse approximation\\
}
\maketitle
\authors{Yu.N.\,Uzikov$^{a,b,c}$\footnote{E-mail: uzikov@jinr.ru} }
\setcounter{footnote}{0}
\from{$^{a}$\, V.P. Dzhelepov Laboratory of Nuclear Problems, Joint Institute for Nuclear Researches, Dubna,  141980 Russia }
%
\from{$^{b}$\, M.V. Lomonosov State University, Physics  Faculty, Moscow, 119991 Russia }
%
%
\from{$^{c}$\,  Moscow Institute of Physics and Technology (National Research University), 141701, Dolgoprudnyi, Moscow region, Russia }


\begin{abstract}

\vspace{0.2cm}

 We have shown that the  vector and tensor  analyzing powers $A_y$, $A_{yy}$
 and also double spin correlation coefficients $C_{y,y}$, $C_{y y,y}$ of  the
 reaction $dd\to n+p+d$  in impulse approximation are linearly connected  to
 corresponding observables of the
 $pd$- elastic scattering. Obtained  relations
  are necessary  for motivation of the polarization experiments for the first  phase of the SPD NICA
 project and for analysis of  expected data.
%
   \end{abstract}
\vspace*{6pt}



\label{sec:intro}
\section*{\bf Introduction}
\label{Uz-Add}

Spin observables of pd-elastic scattering are reasonable well described by  spin dependent
Glauber theory \cite{Platonova:2010zz,Platonova:2010wjt,Temerbayev:2015foa,Platonova:2019yzf}
using   existing experimental data on  spin amplitudes of elastic pp- and pn-scattering.
This means, that  experimental data on spin observables  of the pd-elastic scattering being
compared  with corresponding calculations on the basis of spin-dependent Glauber  theory
can be used as an effective test of  spin amplitudes of pN-elastic scattering
\cite{Uzikov:2020zho}.
Such kind of test  is important at energies of SPD NICA collider
$\sqrt{s_{NN}}= 5-20$ GeV, where  data on
spin pN amplitudes  are very noncomplete and existing phenomenological model parametrizations
for  pN-sccatering \cite{Sibirtsev:2010cjv,Ford:2012dg,Selyugin:2024ccc} need a verification.
Up to now,  Glauber theory of elastic dd-scattering  is widely used for description of unpolarized
differential cross section \cite{Alberi:1970pm}, but   spin-dependent  nucleon-nucleon elastic
scattering amplitudes are
not yet included into  such kind of  formalism. Therefore, the spin observables of dd-elastic scattering cannot
be used for testing spin pN-amplitudes.
Since   only symmetric
colliding mode of dd-collision is planned to  be realised at  SPD NICA, we analyse here the relations between
spin obervables in the
$dd\to n+p+d$ reaction  in the impulse approximation, assuming the mechanism
of the pole diagram in Fig. (\ref{fig1}), and corresponding observables in $pd\to pd$ scattering.
We consider vector and tenzor  analyzing powers $A_y$, $A_{yy}$
 and also spin correlation coefficients $C_{y,y}$, $C_{yy,y}$
 of  the  reaction $dd\to n+p+d$
 and as a result, we show that these observables  are connected with similar  observables
 of the $pd$- elastic scattering
 by linear relations.

\section{\bf  Pole mechanism of  the reaction $dd\to npd$}


\label{Matrfi}
\begin{figure}[t]
\begin{center}
\includegraphics[width=70mm]{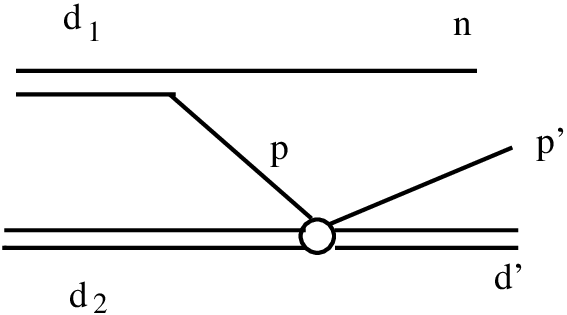}
\vspace{-3mm}
\caption{The pole  mechanism of the reaction $dd\to npd$.}
\end{center}
\labelf{fig1}
\vspace{-5mm}
\end{figure}
\vspace{0.5cm}
Assuming the  pole mechanism (Impulse approximtion)  of the reaction $dd\to npd$ (Fig. \ref{fig1}),
the transition  amplitude can be written as
\begin{eqnarray}
 M_{fi}=\sum _{ \sigma_{p}}
 \frac {M(d_1\to n p){ i}
 M(p d_2\to p'd' ) }
 {p_{p}^2-m^2+i\epsilon},
 \label{Mfi}
\end{eqnarray}
where
$M(p d_2\to pd')$
is the t-matrix of the elastic pd-scattering;
$m$ is the
 mass of the  nucleon;
the product of the nucleon propagator $(p_{p}^2-m^2+i\epsilon)^{-1}$ and
the amplitude of the virtual decay of the deuteron $d_1\to n p$  in Eq. (\ref{Mfi})
 can be written as
\begin{eqnarray}
\frac{ M(d_1\to np)}{({p_{p}}^2-m^2+i\epsilon)}
=
-<\chi_{\sigma_{n}}\chi_{\sigma_{p}}|
\phi_{\lambda_1}({\bf q})>
{{2m}}.
\end{eqnarray}
Here we have the overlap of the deuteron wave function in momentum space for the spin state with
the spin projection $\lambda_1$,
$\phi_{\lambda_1}({\bf q})$, with the Pauli spinors  of the nucleons
$\chi_{\sigma_{n}}$
and $ \chi_{\sigma_{p}}$,
where $\sigma_i$ is
 the {\i}th nucleon spin projection. Then
 the transition matrix element  for the reaction  $dd\to npd$ takes the form
\begin{eqnarray}
 M_{\lambda_1\lambda_2}^{\sigma_{n}\sigma_{p'}\lambda'}
=
K\sum_{ \sigma_{p}}
M_{\sigma_p\lambda_2}^{\sigma_p' \lambda^\prime}(pd\to pd)
<\chi_{\sigma_{n}}\chi_{\sigma_{p}}|\phi_{\lambda_1}({\bf q})>,
\label{me-ddnpd}
\end{eqnarray}
where
\begin{equation}
K=\sqrt{{2m_d}/{4\pi}},
\label{K}
\end {equation}
$K$ is a kinematic factor.
 Assuming dominance of the pole mechnism (Fig. \ref{fig1}), here  we take into account
 only   the S-wave component of the deuteron
 wave function $u(q)$ that gives the following relations:
\begin{eqnarray}
<\chi_{\sigma_{n}}\chi_{\sigma_{p}}|\phi_{\lambda_1}({\bf q})>
=u(q)
(\frac{1}{2}\sigma_{n}\frac{1}{2}\sigma_{p}|1\lambda_1),
\end{eqnarray}
where standard notations for the Clebsh-Gordan coefficients are used.

Squared transition matrix element  (\ref{me-ddnpd}) after summation over spins of final particles and averaging over  spins of initial deuterons
takes the following factorized form
\begin{eqnarray}
 \overline{|M_{\lambda_1\lambda_2}^{\mu_1\mu_2\lambda'}|^2} =
K^2 u^2(q)
\overline{|M_{\lambda_2\mu}^{\lambda'\mu_2}(pd\to pd)|^2}.
\end{eqnarray}
This  factorization of unpolarized cross section is well known for pole mechanism.
A similar factorization for spin observables to our knowledge was
not studied previously and  we do this study in present paper.
Assuming transversal polarizations which is planned at SPD NICA, we consider here
vector ($A_y$) and tenzor $(A_{xx}$)  analyzing powers and  double spin
correlations $C_{y,y}$, $C_{yy,y}$.

To perform this analysis, we introduce the differential cross sections
for initial states with different polarizations of the initial  deuetrons.
 For definite spin projections of
initial deutreons $\lambda_1$  and $\lambda_2$  the  differential cross section of the reaction $dd\to npd$
can be written as
\begin{eqnarray}
d\sigma_{\lambda_1\lambda_2}=K\sum_{\sigma_{n}\sigma_{p'}\lambda'}
| M_{\lambda_1\lambda_2}^{\sigma_{n}\sigma'_{p}\lambda'}(dd\to npd)|^2
\nonumber \\
=K\sum_{\sigma_{n}\sigma_{p'}\lambda'}
|M_{\sigma_p\lambda_2}^{\sigma_p' \lambda^\prime}(pd\to pd)|^2
(\frac{1}{2}\sigma_{n}\frac{1}{2}\sigma_{p}|1\lambda_1)^2u(q_1),
\label{dsiglam1lam2}
\end{eqnarray}
where $\sigma_{p}=\lambda_{1}-\sigma_{n}$.

If the deuteron $d_2$ has the state with  spin projection $\lambda_2$ whereas
the deuteron $d_1$ is not polarized, the corresponding cross
section  of the $d_1d_2\to npd$  reaction is the following
\begin{eqnarray}
d\sigma_{\lambda_2}=\frac{1}{3} \sum_{\lambda_1}\sum_{\sigma_n\sigma_p'\lambda'}
|M_{\lambda_1\lambda_2}^{\sigma_n\sigma_p'\lambda'}(dd\to npd)|^2
=\nonumber \\
=\frac{1}{2} K^2 u^2(q)
\sum_{\sigma_p \lambda'\sigma_p'}
|M_{\lambda_2 \sigma_p}^{\lambda'\sigma_p'}(pd\to pd)|^2.
\label{dslamb2}
\end{eqnarray}
If the deuteron $d_1$ has the state with  spin projection $\lambda_1$ whereas
the deuteron $d_2$ is not polarized, the corresponding cross
section is
\begin{eqnarray}
d\sigma_{\lambda_1}=
\frac{1}{3}\sum_{\lambda_2}\sum_{\sigma_n\sigma_p'\lambda'}
|M_{\lambda_1\lambda_2}^{\sigma_n\sigma_p'\lambda'}(dd\to npd)|^2= \nonumber \\
=\frac{1}{3} K^2 u^2(q)\sum_{\sigma_p\lambda_2}\sum_{\sigma_n\sigma_p'\lambda'}
(\frac{1}{2}\sigma_{n}\frac{1}{2}\sigma_{p}|1\lambda_1)^2
|M_{\lambda_2 \sigma_p}^{\lambda'\sigma_p'}(pd\to pd)|^2.
\label{dslamb1}
\end{eqnarray}

\section{\bf {Vector and tenzor analyzing powers}.}

Vector analyzing power $A_y^{d_2}$,  coressponding to the polarized deuteron $d_2$ in the reaction $dd\to npd$,
is defined as
\begin{eqnarray}
A_y^{d_2}(d_1{ d}^\uparrow_2\to npd) = \frac{d\sigma_{\lambda_2=+1}-d\sigma_{\lambda_2=-1}}
{d\sigma_{\lambda_2=+1}+d\sigma_{\lambda_2=0}+d\sigma_{\lambda_2=-1}}
\label{Ayd2}
\end{eqnarray}
 Using Eq.(\ref{dslamb2})  for $d\sigma_{\lambda_2}$ we find from Eq.(\ref{Ayd2})
 \begin{eqnarray}
 A_y^{d_2}(d_1{ d}^\uparrow_2\to npd)=\nonumber \\
 =\frac{d\sigma_{\lambda=+1}(pd)-d\sigma_{\lambda=-1}(pd)}
{d\sigma_{\lambda=+1}(pd)+d\sigma_{\lambda=0}(pd)+d\sigma_{\lambda=-1}(pd)}\equiv
A_y^d(p{d}^\uparrow\to pd),
\label{Ay-d2}
\end{eqnarray}
 where  $d\sigma_{\lambda}(pd)$ is the differential cross section of the process $pd\to pd$
  for the initial deuteron in the state with spin projection $\lambda$, and $A_y^d(p{\vec d}\to pd)$
  is the vector analyzing power of the process $pd\to pd$  for vector polarized initial deuteron.

Vector analyzing power  $A_y^{d_1}$, corresponding to the polarized deuteron $d_1$, is defined similarly to
Eq.(\ref{Ayd2})  as
\begin{eqnarray}
A_y^{d_1}( {d}^\uparrow_1d_2\to npd) = \frac{d\sigma_{\lambda_1=+1}-d\sigma_{\lambda_1=-1}}
{d\sigma_{\lambda_1=+1}+d\sigma_{\lambda_1=0}+d\sigma_{\lambda_1=-1}}.
\label{Ayd1}
\end{eqnarray}
Using Eq.(\ref{dslamb1}) for the $d\sigma_{\lambda_1}$ we find from Eq.(\ref{Ayd1})
 \begin{eqnarray}
 A_y^{d_1}(d^\uparrow_1 d_2\to npd)= \nonumber \\
 \frac{\sum_{\lambda_2 \lambda'\sigma_p'}\Bigl\{ |M_{\lambda_2,\sigma_p=+1/2}^{\lambda'\sigma_p'}(pd)|^2-
 |M_{\lambda_2,\sigma_p=-1/2}^{\lambda'\sigma_p'}(pd)|^2\Bigr \}}
{\sum_{\lambda_2 \lambda'\sigma_p'}\Bigl\{\frac{3}{2} |M_{\lambda_2,\sigma_p=+1/2}^{\lambda'\sigma_p'}(pd)|^2+
 \frac{3}{2}|M_{\lambda_2,\sigma_p=-1/2}^{\lambda'\sigma_p'}(pd)|^2\Bigr \}}
\equiv
\frac{2}{3}A_y^p({p}^{\uparrow}d\to pd),
\label{Ay-d1}
\end{eqnarray}
where $A_y^p({p}^{\uparrow}d\to pd)$ is the vector analyzing power  of the process $pd\to pd$ in respect
of polarized proton.


The tenzor analyzing power for the polarized  deuteron  $d_2$ is the following
\begin{eqnarray}
A_{yy}(d_1{d}^\uparrow_2\to npd)=
\frac{d\sigma_{\lambda_2=+1}+d\sigma_{\lambda_2=-1} -2 d\sigma_{\lambda_2=0}}
{d\sigma_{\lambda_2=+1}+d\sigma_{\lambda_2=0}+d\sigma_{\lambda_2=-1}}=
A_{yy}(p{ d}^\uparrow\to pd),
\label{Ayy}
\end{eqnarray}
where $A_{yy}(p{d}^\uparrow\to pd)$ is the tenzor analyzing power of the process $pd\to pd$.
Here we used the Eq. (\ref{dslamb2}) for $d\sigma_{\lambda_2}$.

We have found, that the  tenzor analyzing power for the polarized  deuteron  $d_1$,  is equal to zero for
the considered pole mechanims in
the S-wave  approximation for the deuteron.

\section{\bf Double spin correlations}

In order to get  relations between spin correlations $C_{y,y}$ or $C_{yy,y}$  for the $dd\to npd$ and $pd\to pd$ processes
we use  general expressions for spin-dependent differential  cross sections  for collision of two deuterons,
  ${\vec 1}+{\vec 1}$ \cite{Ohlsen:1972zz}
   \begin{eqnarray}
    I=I_0\Bigl [ 1+\frac{3}{2}P_y A_y+\frac{3}{2}P_y^T A_y^T+
    \frac{1}{2}P_yP_{yy}^TC_{y,yy}+\frac{1}{2}P_{yy}P_{y}^TC_{yy,y}+ \nonumber \\
    \frac{9}{4}P_yP_y^TC_{y,y}+\frac{1}{3}P_{yy}A_{yy}+
    \frac{1}{3}P_{yy}^TA_{yy}^T+
    \frac{1}{9}P_{yy}P_{yy}^TC_{yy,yy}\Bigr ].
    \label{ohlsen1}
   \end{eqnarray}
and for  $pd$-collision,
  ${\vec 1}+{\vec {\frac{1}{2}}}$ \cite{Ohlsen:1972zz}:
   \begin{eqnarray}
    I=I_0\Bigl [ 1+\frac{3}{2}P_y A_y+P_y^T A_y^T
    + \frac{1}{3}P_{yy}A_{yy} +
    \frac{3}{2}P_yP_y^TC_{y,y}+\frac{1}{3}P_{yy}P_{y}^TC_{yy,y}
    \Bigr ].
    \label{ohlsen2}
   \end{eqnarray}

   \subsection{\it{Double  spin  vector correlation $C_{y,y}$.}}
\label{subcyy}

When considering the double vector spin correlation $C_{y,y}$, we use the method of Ref. \cite{Uzikov:2023mbh}
applied to the $dd\to npnp$ reaction.
 As follows from Eq. (\ref{ohlsen1}), in order to  get the  double spin correlation $C_{y,y}$,
 one  needs four options for dd-collision:
 (i)  $P_y=\frac{2}{3}$,  $P_y^T=\frac{2}{3}$,
  (ii) $P_y=\frac{2}{3}$, $P_y^T=-\frac{2}{3}$,
  (iii)$P_y=-\frac{2}{3}$,  $P_y^T=\frac{2}{3}$,
  (iv) $P_y=-\frac{2}{3}$, $P_y^T=-\frac{2}{3}$.
 The corresponding event counts  are denoted
 as ${\cal N}_{\uparrow\uparrow}$,  ${\cal N}_{\uparrow\downarrow}$, ${\cal N}_{\downarrow\uparrow}$ and
 ${\cal N}_{\downarrow\downarrow}$ for the first, second, third and fourth options, respectively.
 With these definitions the double spin correlation $C_{y,y}$ is derived from Eq.(\ref{ohlsen1}) as
 \begin{eqnarray}
  C_{y,y}=\frac{(I_{\uparrow\uparrow}-I_{\uparrow\downarrow})+
  (I_{\downarrow\downarrow}-I_{\downarrow\uparrow}) }
  { (I_{\uparrow\uparrow}+I_{\uparrow\downarrow})+
  (I_{\downarrow\downarrow}+I_{\downarrow\uparrow})}.
  \label{cyy-dd}
 \end{eqnarray}
Let us consider the content of the deuteron beam for different polarizations.
Unpolarized  deuteron beam has  components with different  spin-projections on quantization axis
 $\lambda =\pm 1, 0$ in equal portions:
$N_{+}=N_{-}=N_{0}=n$. The spin  ``up'' ($N_{\uparrow}$) and spin ``down'' ($N_{\downarrow}$) beams are
prepared by spin flip of the states $\lambda=1$ and $\lambda=-1$, respectively:
\begin{eqnarray}
N_{\uparrow}=N_{+} + N_{+} +N_0, \\
N_{\downarrow}=N_{-}+N_{-}+N_0.
\label{Nupdown}
\end{eqnarray}
The polarization of the beam  along the axis $OY$ which is directed  along the magnetic field,
is given as the following asymmetry
\begin{eqnarray}
P_{Y}=\frac{N_{\uparrow}-N_{\downarrow}}{N_{\uparrow}+N_{\downarrow}}=\frac{N_{+} - N_{-}}{ N_{+} + N_{0}+N_{-}}.
\label{Py}
\end{eqnarray}
The tensor polarization (alignment) is
\begin{eqnarray}
P_{YY}=\frac{{ N}_{\lambda =+1}+{ N}_{\lambda =-1}-2{ N}_{\lambda =0}}
{{ N}_{\lambda =+1}+{ N}_{\lambda =-1}+{N}_{\lambda =0}}.
\label{Pyy}
\end{eqnarray}
One can see from
Eqs. (\ref{Nupdown}) and (\ref{Py})
that for the deuteron beam content
$N_{\uparrow}$ ($N_{\downarrow}$) the vector polarization of this beam is
$P_y=+\frac{2}{3}$ ($P_y=-\frac{2}{3}$),
 whereas the tensor polarization is zero $P_{YY}=0$ (For experimental realization see, for exmaple, \cite{Stephan:2010zz}).
 Therefore, we conclude  that the considered deuteron beam with $P_{yy}=0$  and  $P_y=\frac{2}{3}$
contains $N_0=n$ deuterons with $\lambda=0$, $N_{+}=2n$ with $\lambda=1$ and no deuterons  with $\lambda=-1$, i.e. $N_{-1}=0$,
whereas the deuteron beam with $P_y=-\frac{2}{3}$ has  $N_0=n$,$N_{+}=0$, $N_{-1}=2n$.


 Introduced in Eq. (\ref{ohlsen1}) counting rates ($I_{\uparrow\uparrow}$, ...)
  can be expressed in relative units (assuming that the luminosity is equal to unit)
  through differential cross sections
 $d\sigma_{\lambda_1\lambda_2}$ defined by Eq. (\ref{dsiglam1lam2}):
 \begin{eqnarray}
 {\cal N}_{\uparrow\uparrow}= 2\cdot 2d\sigma_{++}+2d\sigma_{+0}+ 2d\sigma_{0+}+d\sigma_{00},\nonumber \\
{\cal N}_{\uparrow\downarrow}=2\cdot 2d\sigma_{+-}+ 2d\sigma_{0-}+2d\sigma_{+0}+  d\sigma_{00},\nonumber \\
{\cal N}_{\downarrow\uparrow}=2\cdot 2d\sigma_{-+}+2d\sigma_{0+}+ 2d\sigma_{-0}+d\sigma_{00},\nonumber \\
{\cal N}_{\downarrow\downarrow}= 2\cdot 2d\sigma_{--}+2d\sigma_{-0}+ 2d\sigma_{0-}+d\sigma_{00}.
\label{Nupdown}
\end{eqnarray}
Here we have taken into account  the above considered relations between beam polarizations $P_y=\pm \frac{2}{3}$ and
relative numbers of the deuterons
$N_0$, $N_{+}$, $N_{-1}$ in the beam.

  Using   Eqs. (\ref{dsiglam1lam2}) one  can calculate nine differential cross sections $d\sigma_{\lambda_1\lambda_2}$
  entering Eq. (\ref{Nupdown}) and then find the double spin correlation coefficient as
  \begin{eqnarray}
  C_{y,y}(dd\to pnd)= \nonumber \\
  \frac{\frac{2}{3}\Bigl (|M_{+1,+\frac{1}{2}}|^2+|M_{-1,-\frac{1}{2}}|^2-|M_{+1,-\frac{1}{2}}|^2-
  |M_{-1,+\frac{1}{2}}|^2 \Bigr )}
  {|M_{+1,+\frac{1}{2}}|^2+|M_{-1,-\frac{1}{2}}|^2+ |M_{0,+\frac{1}{2}}|^2+|M_{0,-\frac{1}{2}}|^2+|M_{+1,-\frac{1}{2}}|^2+|M_{-1,+\frac{1}{2}}|^2},
  \label{cyy-dd-m}
  \end{eqnarray}
where $|M_{\lambda,\sigma_p}|^2= \sum_{\lambda,\sigma_p}|M_{\lambda,\sigma_p}^{\sigma_p'\lambda'}(pd\to pd)|^2$ is
the
transition matrix element of the process $pd\to pd$ squared  and summed over  spins of the final particles.
  To find the double spin correlation  $C_{y,y}(pd\to pd)$ for the $pd\to pd$ we use Eq. (\ref{ohlsen2}) considering
  four combinations
  of polarizations  of initial proton $P_y=\pm1$ and deuteron $P_y^T=\pm \frac{2}{3}$:
  \begin{eqnarray}
 {I }_{\uparrow\uparrow}(P_y=1,P_y^T=\frac{2}{3}) = I_0(1+\frac{3}{2}A_y+\frac{2}{3}A_y^T+C{y,y}),\nonumber \\
 {I}_{\downarrow\uparrow}(P_y=-1,P_y^T=\frac{2}{3})=  I_0(1-\frac{3}{2}A_y+\frac{2}{3}A_y^T-C{y,y}),\nonumber \\
{I}_{\uparrow\downarrow}(P_y=1,P_y^T=-\frac{2}{3})= I_0(1+\frac{3}{2}A_y-\frac{2}{3}A_y^T-C{y,y}),\nonumber \\
{I}_{\downarrow\downarrow}(P_y=-1,P_y^T=-\frac{2}{3})= I_0(1-\frac{3}{2}A_y-\frac{2}{3}A_y^T+C{y,y}).
\label{ipd-updown}
\end{eqnarray}
These relations  lead to the double spin correlation $C_{y,y}(pd\to pd)$ in terms of differential
$pd$-elastic scatterring cross sections  $I(P_y, P_y^T)$, given by Eq. (\ref{ohlsen2}) in the form
which exactly coincides with Eq.(\ref{cyy-dd}).
On the other hand, $I(P_y, P_y^T)$ can be expressed via differential cross sections of pd-elastic scattering
for definite values of spin projections $\sigma_p$ and  $\lambda$,
$d\sigma_{\sigma_p,\lambda}(pd)$,
as
  \begin{eqnarray}
 {I }_{\uparrow\uparrow}(P_y=1,P_y^T=\frac{2}{3}) = 2d\sigma_{\frac{1}{2},1}+ d\sigma_{\frac{1}{2},0},\nonumber \\
 {I}_{\downarrow\uparrow}(P_y=-1,P_y^T=\frac{2}{3})= 2d\sigma_{-\frac{1}{2},1}+ d\sigma_{-\frac{1}{2},0},\nonumber \\
{I}_{\uparrow\downarrow}(P_y=1,P_y^T=-\frac{2}{3})= 2d\sigma_{\frac{1}{2},-1}+ d\sigma_{\frac{1}{2},0},\nonumber \\
{I}_{\downarrow\downarrow}(P_y=-1,P_y^T=-\frac{2}{3})= 2d\sigma_{-\frac{1}{2},-1}+ d\sigma_{-\frac{1}{2},0}.
\label{ipd-updown-sigm-lamb}
\end{eqnarray}
Here we have taken into account that the considered deuteron beam with $P_{yy}=0$  and  $P_y=\frac{2}{3}$
 has  the "ocupation" numbers $N_0=n$,$N_{+}=0$, $N_{-1}=2n$,
whereas the deuteron beam with $P_y=-\frac{2}{3}$ has  $N_0=n$,$N_{+}=0$, $N_{-1}=2n$.
Substituting these expressions into the formula for  $C_{y,y}(pd\to pd)$, which as was shown above,
is given by Eq.(\ref{cyy-dd}),
we find that the right hand of Eq.(\ref{cyy-dd-m})  coincides with $C_{y,y}(pd\to pd)$ with the factor $\frac{2}{3}$:
 \begin{eqnarray}
 C_{y,y}(dd\to npd)=\frac{2}{3}C_{y,y}(pd\to pd)
 \label{cyy-dd-pd}
\end{eqnarray}


\subsection{\it{Double tensor-vector correlation $C_{yy,y}$.}}
\label{subcyyy}
 To get  tenzor-vector dd-correlation  $C_{yy,y}$ we use four options:  $P_y=0, P_{yy}=\pm 1$ for
 the deuteron $d_1$ and
 $P_y=\pm \frac{2}{3}, P_{yy}=0$ for the deuteron  $d_2$.
 Using this options   one  can  find the coefficient $C_{yy,y}$ from Eq.(\ref{ohlsen1}) in the following form
 \begin{eqnarray}
   C_{yy,y}(dd)=3\frac{(I_{+\uparrow}-I_{+\downarrow})+
  (I_{-\downarrow}-I_{-\uparrow}) }
  { (I_{+\uparrow}+I_{+\downarrow})+
  (I_{-\downarrow}+I_{-\uparrow})},
  \label{cyy,y}
 \end{eqnarray}
where $I_{d_1 d_2}=I_{+\uparrow}$ is the counting rate for dd-collision with $P_{yy}=+1$ for the deuteron $d_1$
and $P_y=\frac{2}{3}$ for the deuteron $d_2$, and so on.

The occupation numbers $N_{\pm 1}, N_0$ for the beam $d_2$  were given in the previous subsection \ref{subcyy}.
For the deuteron beam $d_1$ (with $P_y=0$)
we find  from Eqs. (\ref{Py},\ref{Pyy})  $N_{+}=\frac{3}{2}n$, $N_{-}=\frac{3}{2}n$, $N_0=0$ for $P_{yy}=+1$,
and $N_{+}=\frac{1}{2}n$, $N_{-}=\frac{1}{2}n$, $N_0=2n$ for $P_{yy}=-1$.
Counting rates $I_{\pm,\updownarrow}$ entering Eq. (\ref{cyy,y}) can be expressed in terms of differential cross sections
$d\sigma_{\lambda_2\lambda_1}$, defined by Eq.(\ref{dsiglam1lam2}), in the following way:
\begin{eqnarray}
I_{+\uparrow}=3 d\sigma_{++}+3d\sigma_{-+}+\frac{3}{2}d\sigma_{+0}+\frac{3}{2}d\sigma_{-0}, \nonumber \\
I_{-\uparrow}= d\sigma_{++}+ d\sigma_{-+}+ 4d\sigma_{0+} +\frac{1}{2}d\sigma_{+0} +\frac{1}{2}d\sigma_{-0}
  +2d\sigma_{00}, \nonumber \\
I_{+\downarrow}=3 d\sigma_{+-}+3d\sigma_{--}+\frac{3}{2}d\sigma_{+0}+\frac{3}{2}d\sigma_{-0}, \nonumber \\
I_{-\downarrow}=d\sigma_{+-}+ d\sigma_{--}+ 4d\sigma_{0-} +\frac{1}{2}d\sigma_{+0} +\frac{1}{2}d\sigma_{-0}
  +2d\sigma_{00}.
\label{Icyyy}
\end{eqnarray}
Using  Eq.(\ref{dsiglam1lam2})  for $d\sigma_{\lambda_2\lambda_1}$, we find from Eqs.(\ref{cyy,y},\ref{Icyyy})
\begin{eqnarray}
C_{yy,y}(dd)=\frac{3}{2}\frac{(d\sigma_{+,\frac{1}{2}}-d\sigma_{+,-\frac{1}{2}})+
(d\sigma_{-,\frac{1}{2}}-d\sigma_{-,-\frac{1}{2}})-2(d\sigma_{0,\frac{1}{2}}-d\sigma_{0,-\frac{1}{2}})}
{\sum_{\lambda,\sigma_p} d\sigma_{\lambda,\sigma_p}  }.
 \label{cyy,y-dd-f}
 \end{eqnarray}

Now we consider the $C_{yy,y}(pd)$ correlation in the  process $pd\to pd$.
Assuming   for the deuteron beam  polarisations $P_y^T=0$, $P_{yy}^T=\pm 1$ and for the proton beam --
$p_y^p=\pm 1$, we find from Eq. (\ref{ohlsen2})
 \begin{eqnarray}
   C_{yy,y}(pd)=3\frac{(I^{pd}_{+\uparrow}-I^{pd}_{+\downarrow})+
  (I^{pd}_{-\downarrow}-I^{pd}_{-\uparrow}) }
  { (I^{pd}_{+\uparrow}+I^{pd}_{+\downarrow})+
  (I^{pd}_{-\downarrow}+I^{pd}_{-\uparrow})},
  \label{cyy,y-pd}
 \end{eqnarray}
where $I^{pd}_{P_y^T, p_y^p}$ is the  counting rate for given values of $P_y^T$ and $p_y^p$ which can be expressed via
the differential cross sections of pd-scattering  $d\sigma_{\lambda,\sigma_p}$:
\begin{eqnarray}
I^{pd}_{+\uparrow}= \frac{3}{2}d\sigma_{+,\frac{1}{2}}+\frac{3}{2}(d\sigma_{-,\frac{1}{2}}), \nonumber \\
I^{pd}_{+\downarrow}= \frac{3}{2}d\sigma_{+,\frac{1}{2}}+\frac{3}{2}d\sigma_{-,-\frac{1}{2}}, \nonumber \\
 I^{pd}_{-\uparrow}  =\frac{1}{2}d\sigma_{+,\frac{1}{2}}+\frac{1}{2}d\sigma_{-,\frac{1}{2}} +2d\sigma_{0,\frac{1}{2}}, \nonumber \\
I^{pd}_{-\downarrow}=\frac{1}{2}d\sigma_{+,-\frac{1}{2}}+\frac{1}{2}d\sigma_{-,-\frac{1}{2}} +2d\sigma_{0,-\frac{1}{2}}.
  \label{I-pd}
 \end{eqnarray}
Using these relations, one can find from Eq.(\ref{cyy,y-pd}):
\begin{eqnarray}
C_{yy,y}(pd)=\frac{3}{2}\frac{(d\sigma_{+,\frac{1}{2}}-d\sigma_{+,-\frac{1}{2}})+
(d\sigma_{-,\frac{1}{2}}-d\sigma_{-,-\frac{1}{2}})-2(d\sigma_{0,\frac{1}{2}}-d\sigma_{0,-\frac{1}{2}})}
{\sum_{\lambda,\sigma_p} d\sigma_{\lambda,\sigma_p}  }.
 \label{cyy,y-pd-f}
 \end{eqnarray}
that exactly coincides with Eq.(\ref{cyy,y-dd-f}). Therefore, finally we have found the following relation
\begin{eqnarray}
C_{yy,y}(dd\to npd )=C_{yy,y}(pd\to pd).
\label{cyy,y-dd-pd}
\end{eqnarray}


 \section*{\bf Summary}

  The main results of this work are given  in Eqs. (\ref{Ay-d2}),(\ref{Ay-d1}),(\ref{Ayy}) for vector and tenzor analyzing powers
 and in Eqs. (\ref{cyy-dd-pd}),
 (\ref{cyy,y-dd-pd}) for double spin correlations.
The obtained results are restricted by the pole diagram for the transition amplitutde of the reaction $dd\to npd$
 and S-wave approximation for the deuteron wave function. Both this approximations are expected to be valid at small internal relative
 momenta $q<0.15 GeV/c$ between nucleons in the deuteron. At higher momenta $q>0.2$GeV/c the contribution of the D-wave
 increases and has to be taken into accout.
 Furthermore, at momenta  $q>0.2$GeV/c, the contribution of rescatterings of final  proton and deuteron on the final neutron is
 increasing because the relative distances between nucleons in the deuteron $d_1$ become smaller.
 These effects will be considered in a separate publication.

{\bf Acknowledgements.}
The research was carried out at the expense of the grant of the Russian Science Foundation No. 25-72-30005, https://rscf.ru/project/25-72-30005/

\bibliographystyle{pepan}
\bibliography{pepan_biblio-uzdd}

\end{document}